\documentclass[aps,twocolumn,pra,amssymb,floatfix,superscriptaddress,10pt]{revtex4}

\usepackage{graphicx}
\usepackage{mathptmx}

\begin{document}

\title{Parametric resonance of capillary waves at the interface between two immiscible Bose-Einstein condensates}

\author{D. Kobyakov}
\affiliation{Department of Physics, Ume{\aa} University, 901 87 Ume{\aa}, Sweden}

\author{V. Bychkov}
\affiliation{Department of Physics, Ume{\aa} University, 901 87 Ume{\aa}, Sweden}

\author{E. Lundh}
\affiliation{Department of Physics, Ume{\aa} University, 901 87 Ume{\aa}, Sweden}

\author{A. Bezett}
\affiliation{Institute for Theoretical Physics, Utrecht University, Leuvanlaan 4, 3584 CE Utrecht, The Netherlands}

\author{M. Marklund}
\affiliation{Department of Physics, Ume{\aa} University, 901 87 Ume{\aa}, Sweden}

\begin{abstract}
We study parametric resonance of capillary waves on the interface between two immiscible Bose-Einstein condensates pushed towards each other by an oscillating force. Guided by analytical models, we solve numerically the coupled Gross-Pitaevskii equations for two-component Bose-Einstein condensate at zero temperature.
We show that, at moderate amplitudes of the driving force, the instability is stabilized due to non-linear modifications of the oscillation frequency. When the amplitude of the driving force is large enough, we observe  detachment of droplets from the Bose-Einstein condensates, resulting in generation of quantum vortices (skyrmions). We analytically investigate the vortex dynamics, and conditions of quantized vortex generation.
\end{abstract}

\maketitle

\section{Introduction}
Numerous recent studies of hydrodynamic phenomena in quantum media have demonstrated remarkable interplay between quasi-classical hydrodynamics and purely quantum effects such as quantum solitons and vortices, e.g. see \cite{Becker-et-al-solitons-NatPhys,Hamner-et-al-counterflow-inst,Nardin-et-al-vortices-NatPhys,Blaaugeers-et-al-KH-exper,Takeuchi-2010-KH,Suzuki_KH_BECs}. As an example of such interplay, we can mention works on shock fronts in Bose-Einstein condensates (BECs), non-linear optics and quantum plasmas \cite{Wan-et-al-NatPhys-2007,Hoefer-et-al-2006,Bychkov-et-al-2008}: the quantum shocks propagate as soliton trains due to  Bohm-de Broglie dispersion instead of the monotonic transition between low-density and high-density  gases (fluids, plasmas) in a classical shock.
Development of quantum solitons has been also obtained experimentally and theoretically in the process of dynamical quantum interpenetration of both miscible and immiscible BECs
\cite{Hamner-et-al-counterflow-inst,Hoefer-et-al-MI-solitons-2011,Kobyakov_QuantumSwapping-2012}.

Another important example of the interplay concerns generation of quantized vortices. Traditionally, quantum vortices in BECs are produced by rotating the trap \cite{Bretin-et-al-2004-trap-rot}, by stirring  BECs by moving potentials
\cite{Sasaki-2010-vortices}, by coherent transfer of the orbital angular momentum of photons to atoms
\cite{Andersen-2006-Laguerre-Gaussian-to-bec-vortex}, by adiabatic phase imprinting
\cite{VortexPumpInBECs-Mottonen-et-al}, or by modulational instability of solitons
\cite{soliton_to_vortex_exp,soliton_to_vortex_theor}.
Recently, there has been also much interest in multi-component BECs with a well-distinguished interface, which allow the possibility of vortex generation by means of quasi-classical hydrodynamic instabilities: the Kelvin-Helmholtz instability, the Rayleigh-Taylor (RT) instability, the capillary instability, etc. \cite{Blaaugeers-et-al-KH-exper,Takeuchi-2010-KH,Suzuki_KH_BECs,Sasaki-2009-RT,Bezett-et-al-RM,Sasaki-2011-capillary,Kobyakov-2011-linear}.
With the help of the instabilities, one can produce rather complicated quantum vortex structures in BECs like skyrmions, for which the intrinsically empty vortex core in one BEC-component is filled by the other component
\cite{Sasaki-2011-capillary,Kobyakov_QuantumSwapping-2012}.
The system of a two-component  BEC can be realized experimentally by blocking the spin-recombination process $|1,-1\rangle\,+\, |1,1\rangle\rightarrow|1,0\rangle\,+\, |1,0\rangle$ by the quadratic Zeeman effect, as has been discussed in
\cite{Sasaki-2009-RT}. In the case of magnetic gradient pushing the BEC components to each other, they tend to  reduce the potential energy of the system by exchanging places, which typically happens in the form of a multidimensional flow due to the RT instability. In that sense, the magnetic field gradient plays the role similar to a gravitational field for the classical RT instability.
As an alternative to the RT instability, Ref.
\cite{Kobyakov_QuantumSwapping-2012} has  found the possibility of reducing the excess of potential energy by 1D dynamical quantum interpenetration; still, it has been also shown that the RT instability dominates for interface perturbations of a sufficiently large wavelength.

Linear stability analysis based on the variational principle
\cite{Kobyakov-2011-linear} has  demonstrated that the same experimental configuration may also reproduce other quasi-hydrodynamic phenomena in BECs such as the Richtmyer-Meshkov instability and the parametric instability by using time-dependent magnetic field, which corresponds to time-dependent effective "gravity". We focus on the parametric resonance, which is also known as the parametric or Faraday instability. We point out that interface dynamics driven by time-dependent "gravity" is intensively studied within the classical fluid (gas) mechanics, e.g. see recent numerical simulations of the Richtmyer-Meshkov instability in gases
\cite{Bai2010-RM,Dimonte2010-RM,Hahn-2011-RM},  laboratory experiments on fluids with oscillating effective "gravity" \cite{DimonteSchneider1996,Amiroudine-2012-Faraday-exp}, works on inertial confined fusion
\cite{Betti1993,Kawata1993}
and combustion
\cite{Searby1992,Bychkov-1999,Bychkov2005,Bychkov-2008-DDT}.
In the classical hydrodynamics, the parametric instability is typically excited at the interface between  light and heavy gases (fluids) with the "gravity" produced by direct vibrations of the experimental set-up or by acoustic waves (the later is especially typical for combustion systems). In particular, experimental and numerical studies of the parametric instability in combustion encountered powerful turbulence produced in the flow because of the instability \cite{Searby1992,Petchenko2006,Petchenko2007}.

In the present paper we study development of the parametric instability at the interface of two immiscible BEC components pushed towards each other by an oscillating force. The instability arises due to the parametric resonance pumping quantum capillary waves at the interface. We show that in the present configuration the instability does not lead to turbulence. At moderate amplitudes of the driving force the instability is stabilized at the nonlinear stage due to modifications of the doubled oscillation frequency in comparison with frequency of the driving force. When the amplitude of the driving force is large enough, we observe  detachment of droplets from BEC components and generation of quantum vortices (skyrmions). We discuss properties and dynamics of the skyrmions.

\section{Analytical model for the parametric resonance}

Two immiscible BECs are separated in space in the ground state of the system. To investigate physics of the dynamical system it is sufficient to study a symmetric case, where the components' atoms have equal masses, and equal intra-component interaction parameters $g\equiv g_{11}=g_{22}$,  $g_{12}>g$, where $g_{ij}\equiv 4\pi\hbar^2a_{ij}/m$ and $a_{ij}$ are the scattering lengths for collisions between atoms of the $i$-th and $j$-th components. The scattering lengths are calculated depending on the geometry of system confinement
\cite{bib_PethickSmith}; the dimensionless parameter $\gamma\equiv g_{12}/g-1>0$ specifies strength of mutual repulsion of the components.
We study a ribbon-shaped 2D geometry where the BEC components are initially placed in the domains $y<0$ and $y>0$, respectively. The BEC is tightly confined in the $z$ direction; in the $y$ direction it is trapped so that the Thomas-Fermi (TF) approximation for each component holds well
\cite{Kobyakov_QuantumSwapping-2012}; finally, it is not confined in the $x$ direction.  The system of width ${\tilde \lambda}$ along the $x$ direction is shown in Fig. 1, with periodic boundary conditions imposed along the $x$-axis. Within the analytical models the ribbon may be treated as initially uniform in the $y$ direction with negligible influence of the trap edges. Still, finite length of the ribbon is required in the numerical solution to the Gross-Pitaevskii equations.

In the case of zero external forcing, linear stability analysis \cite{Waves_on_the_interface_between_two_BECs} predicts a perturbation mode,
which may be interpreted as quantum capillary waves with the frequency ${{\rm{\tilde \Omega }}_c}$  depending on the wave number ${\tilde k = 2\pi / \tilde \lambda}$ as
\begin{equation}
\label{capillary}
{{\rm{\tilde \Omega }}_c^{2}} = {\tilde k^3}{{{\hbar ^2}\tilde n_0^{1/2}} \over {2{m^2}}}\sqrt {2\pi \left( {{a_{12}} - a} \right)},
\end{equation}
where tilde indicates dimensional values and $\tilde n_0$ is the condensate density at the unperturbed infinitely thin interface (i.e. in the Thomas-Fermi approximation). Taking into account a time-dependent  magnetic force pushing BECs to each other, we obtain an equation for the linear  interface perturbations \cite{Kobyakov-2011-linear}
\begin{equation}
\label{eq1.00}
\left[ \frac{{{d}^{2}}}{d{{\tilde{t}}^{2}}}+{{\rm{\tilde \Omega }}_c^{2}}-\tilde{k}\frac{{{\mu }_{B}}{{B}^{'}}( \tilde{t} )}{2m} \right]\tilde{\zeta} ( \tilde{t} )=0,
\end{equation}
where $\tilde{\zeta} ( \tilde{t} )$ is amplitude of deviation of interface from hydrostatic equilibrium, $\mu_B$ is the Bohr magneton and $B'(\tilde{t})$ indicates the external (Stern-Gerlach) force due to the magnetic field gradient.
\begin{figure}
\includegraphics[width=3.6in]{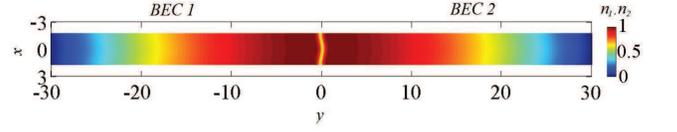}
\caption{(Color online) The system geometry: density distribution for two immiscible BECs, $n_1$ and $n_2$, for $R_0=30$, with initial perturbations at the interface. The BECs are tightly trapped in $z$ direction (means quasi-2D geometry), trapped with resulting TF profile in $y$ direction, and is uniform in $x$ direction. Dimensionless units are defined in Sec II.}
\end{figure}

We will consider a harmonically oscillating force, which may lead to the parametric instability.
For simplicity of designations, we use the standard scalings for a trapped system with the length scales measured in units of the oscillator characteristic length $a_y=\sqrt{\hbar/m\omega_y}$, time in units of $(2\omega_y)^{-1}$, and wave functions in units of $\sqrt{\tilde{n}_0}$.
Within these designations, the dimensionless system size is ${{R}_{0}}=\sqrt{2g\tilde{n}_0/\hbar \omega_y}$ and the dispersion relation for the capillary waves is $\Omega _{c}^{2}=\sqrt{\gamma} {{R}_{0}}{{k}^{3}}$.
In the same manner we define the dimensionless external force $b(t)={{\mu }_{B}}{B'(t)}{{a}_{y}}/\hbar {{\omega }_{y}}$, which we take in the form
\begin{equation}
\label{eq1.02}
b(t)=b_c \sin(\Omega t).
\end{equation}
The approximate analytical solution to Eq. (\ref{eq1.00}) with the oscillating force may be obtained using  the method of \cite{LL-Mechanics}, which describes exponential growth of interface perturbations as
\begin{equation}
\label{perturbations}
\zeta  = \zeta _0 \sin \left( \Omega t/2+\varphi \right) \exp \left( \alpha t \right)
\end{equation}
with the growth rate $\alpha$, some amplitude ${\zeta _0}$, and the phase shift $\varphi$ with respect to the driving force.
Mark that the growth of the parametric instability is accompanied by interface oscillations with frequency ${\rm{\Omega }}/2$. Then the  perturbation growth rate may be found analytically using the equation
\begin{equation}
\label{growth-rate}
{\left( {{\alpha ^2} - {{{{\rm{\Omega }}^2}} \over 4} + {\rm{\Omega }}_{\rm{c}}^2} \right)^2} - {{b_c^2{k^2}} \over 4} + {\alpha ^2}{{\rm{\Omega }}^2} = 0,
\end{equation}
see \cite{Kobyakov-2011-linear} for details; the growth rate is plotted vs the perturbation wave number in Fig. 2 for $\Omega = 8.55$ and different amplitudes of the driving force.  As we can see, maximal growth rate is achieved for the frequency of the driving force equal doubled capillary frequency, ${\rm{\Omega }}=2{\rm{\Omega }}_{c}$, which is the condition of a parametric resonance. In that case the growth rate is given by a simple formula
\begin{equation}
\label{eq1.03}
{{\alpha }^{2}}=\sqrt{4 \Omega _{c}^{4}+{{ {{b}_{c}^{2}}k^{2} }}}-2 \Omega _{c}^{2},
\end{equation}
and the parametric instability may be excited even by an extremely weak force. For other frequencies of the driving force, out of the resonance ${\rm{\Omega }}\neq 2{\rm{\Omega }}_{c}$, a finite force amplitude $b_c$ is required to produce the instability.
\begin{figure}
\includegraphics[width=3.6in]{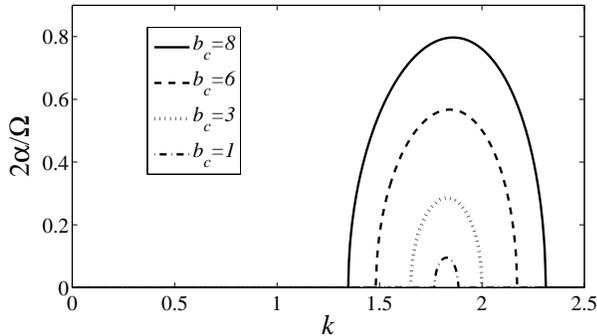}
\caption{Scaled growth rate of the parametric instability vs the wave number of the interface perturbation between BECs for $\Omega = 8.55$ and different amplitudes of the driving force, obtained in \cite{Kobyakov-2011-linear}. Dimensionless units are defined in Sec II.}
\end{figure}
We point out that in the case of moderate amplitude of the driving force, the instability domain in wave numbers is strongly localized around the resonance point, see Fig. 2. Another interesting characteristic of the linear stage is the relative value of the instability growth rate $\alpha$ and the perturbation frequency $\Omega/2$ in Eq. (\ref{perturbations}), which shows how fast the perturbations grow in one oscillation period of the driving force. As we will see below, the instability parameter $2\alpha/\Omega$ determines strength of the process and qualitative regime of the parametric resonance at the nonlinear stage.

Let us qualitatively discuss dynamics of the interface expected at the nonlinear stage of the parametric resonance, for the cases of weak instability ($2\alpha/\Omega \ll 1$) and strong instability ($2\alpha/\Omega \gg 1$). The transition between these two regimes happens for $2\alpha/\Omega \sim 1$. In the case of weak instability one should expect nonlinear stabilization of the parametric resonance because of nonlinear modifications of the oscillation frequency \cite{LL-Mechanics}. Taking into account small but finite amplitude of the oscillations $\zeta_{0}$, one finds the oscillation frequency modified from $\Omega/2$ to $\Omega/2-\kappa \zeta_{0}^{2}$, where $\kappa$ is some factor, which may be found, e.g., from the numerical solution. Since the frequency changes together with the growth of the perturbation amplitude, then the system eventually leaves the instability domain and the perturbation growth is saturated.
Substituting the modified frequency into Eq. (\ref{growth-rate}) and taking $\alpha = 0$ we evaluate the maximal oscillation amplitude attained in the resonance as
\begin{equation}
\label{maximal-amp}
\zeta_{0}=\sqrt{{b}_{c}k/2 \Omega \kappa}.
\end{equation}
In the opposite limit of strong instability, $2\alpha/\Omega \gg 1$, the approach of small corrections to the linear solution employed in \cite{LL-Mechanics} fails. In that case we expect considerable growth of the perturbation amplitude already on one  oscillation period with strongly nonlinear effects resembling qualitatively the RT instability, though with time-dependent effective gravity.

\section{Numerical solution}
We investigate the parametric resonance numerically  by solving the coupled Gross-Pitaevskii equations
\begin{eqnarray}
\label{GP1}
i\hbar \frac{\partial }{\partial \tilde{t}}{{\tilde{\psi} }_{1}}=\left[ -\frac{{{\hbar }^{2}}}{2{{m}}}\Delta_{x,y}+{{\tilde{V}}_{1}}+{{g}_{11}}{{\left| {{\tilde{\psi}}_{1}} \right|}^{2}}+{{g}_{12}}{{\left| {{\tilde{\psi}}_{2}} \right|}^{2}} \right]{{\tilde{\psi} }_{1}}, \\
\label{GP2}
i\hbar \frac{\partial }{\partial \tilde{t}}{{\tilde{\psi} }_{2}}=\left[ -\frac{{{\hbar }^{2}}}{2{{m}}}\Delta_{x,y}+{{\tilde{V}}_{2}}+{{g}_{22}}{{\left| {{\tilde{\psi}}_{2}} \right|}^{2}}+{{g}_{12}}{{\left| {{\tilde{\psi} }_{1}} \right|}^{2}} \right]{{\tilde{\psi}}_{2}},
\end{eqnarray}
where $\Delta_{x,y}\equiv\partial_{xx}^2+\partial_{yy}^2$, and the potentials ${{\tilde{V}}_{1,2}}(t,x,y)$ include both the trapping potential and the driving potential as ${{\tilde{V}}_{j}}\equiv m\omega^2\left(x^2+y^2\right)/2+(-1)^j \mu_B B'(t)y/2\equiv\tilde{V}_{t}+(-1)^j \tilde{V}_{d}$. The trapping potential along $z$-axis is supposed to be much stronger than the two-particle interaction energy, and this leads to renormalization of the interaction parameters \cite{bib_PethickSmith}. In all calculations, we keep the scaled system size $R_0=30$ and the repulsion parameter $\gamma = 0.01$, which implies a sufficiently small thickness  of the interface $\sim 1/\sqrt{\gamma }R_{0} \approx 0.3$ and a much smaller healing length $\sim 1/R_{0} \approx 3.3 \cdot 10^{-2}$ \cite{Kobyakov_QuantumSwapping-2012}. The unperturbed planar distribution of BEC density has been obtained numerically similar to \cite{Bezett-et-al-RM,Kobyakov_QuantumSwapping-2012}.

At $t=0$ we impose a cosine-shaped  perturbation at the interface as shown in Fig. 1, which corresponds to the capillary mode with the perturbation wave length $\lambda=3.44$, and immediately turn on the driving magnetic (Stern-Gerlach) force, and study dynamics of the system for different amplitudes of the force $b_{c}=1-10$. Since we are interested in the parametric resonance, then we chose the external force frequency $\Omega=8.55$ equal to $2\Omega_{c}$ as calculated for the wavelength $\lambda=3.44$ of the initial perturbations.
In order to define position of the interface in the study, we chose  the  density level  equal to the density of  the unperturbed solution at $z=0$.
\begin{figure}
\includegraphics[width=3.6in]{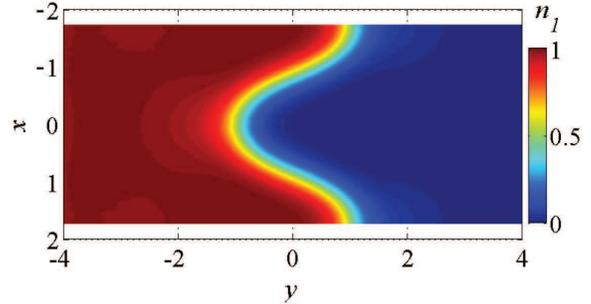}
\caption{(Color online) Density distribution of BEC 1, $n_1$, for $R_0=30$, $b_c=3$ at $t=2.0$, showing the maximal interface deformation in the regime of non-linearly stabilized parametric resonance. Dimensionless units are defined in Sec II.}
\end{figure}
\begin{figure}
\includegraphics[width=3.6in,height=2.2in]{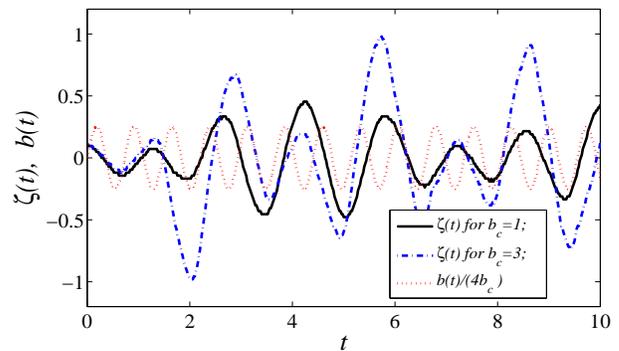}
\caption{(Color online) Amplitude $\zeta(t)$ of the interface deviation from equilibrium  for $R_0=30$ and $b_c=1,\, 3$ and the scaled amplitude of the driving force (dots). Dimensionless units are defined in Sec II.}
\end{figure}
\begin{figure}
\includegraphics[width=3.6in,height=2.2in]{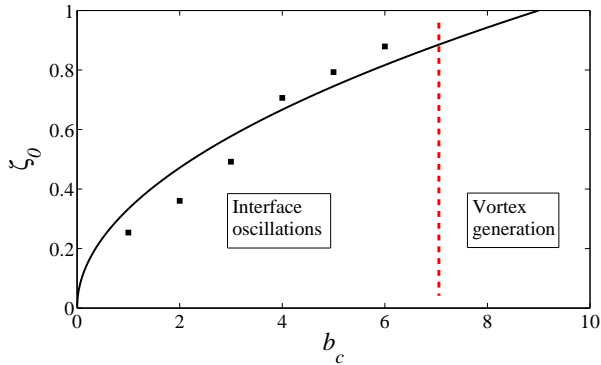}
\caption{(Color online) Phase diagram of vortex generation in the fixed-size (fixed $R_0$) system, showing possible regimes with well-defined interface. For $b_c\geq10$ the system is not well-described in the mean-field approximation because it acquires too much incoherent excitations. Mean deviation of the interface from equilibrium $\zeta_0$, averaged on twice as many periods of modulations, as shown in Fig. 4, for various $b_c$ (time-averaged for irregular dependence) $R_0=30$. The solid line corresponds to the analytical model (\ref{maximal-amp}) with $\kappa = 0.96$. The dashed line separates two qualitatively different nonlinear regimes. Dimensionless units are defined in Sec II.}
\end{figure}
\begin{figure}
\includegraphics[width=3.6in]{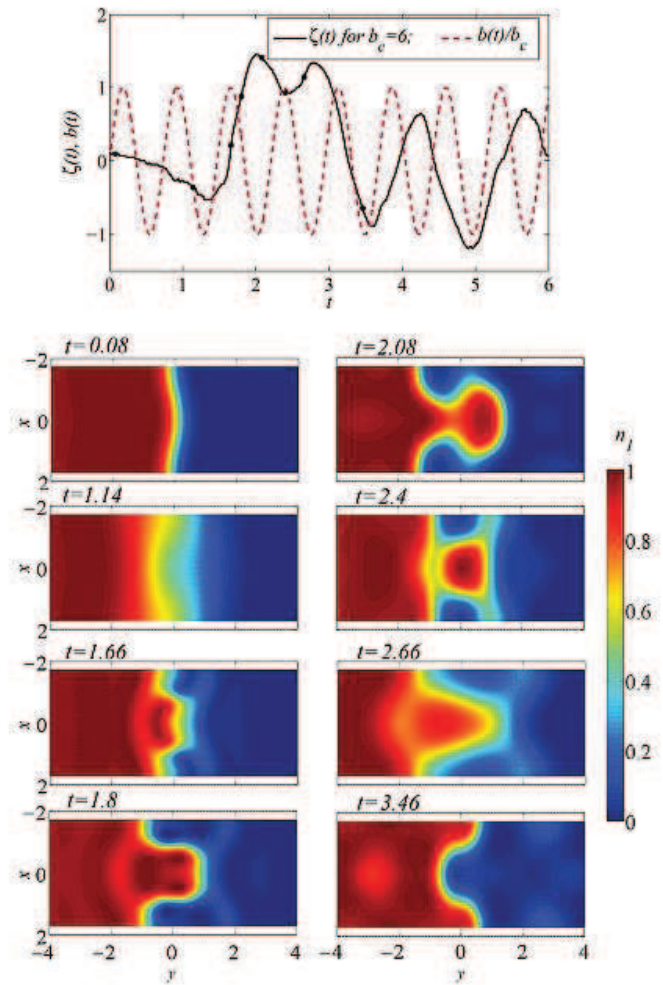}
\caption{(Color online) (Above) The scaled amplitude of the driving force, and amplitude of the interface deviation from equilibrium  for $R_0=30$ and $b_c=6$; the markers indicate the time instants for the density snapshots below. (Below) Snapshots of density distribution of BEC 1 $n_1$. Dimensionless units are defined in Sec II.}
\end{figure}

System dynamics at the nonlinear stage of the resonance depends qualitatively on the magnitude of the external force $b_c$. In the case of relatively weak force, e.g. for $b_c=1,\, 3$, the shape of the interface resembles the initial cosine-shaped perturbations, though with increased amplitude. As an example, maximal distortion of the interface for $b_c=3$ is demonstrated in Fig. 3  for the time instant $t = 2.0$. Time dependence of the perturbation amplitude for $b_c=1,\, 3$ may be characterized as rather regular modulated oscillations, see Fig. 4, with frequency $\approx \Omega/2$ and with the maximal interface distortion attained after 3-4 oscillations. After that, dynamics of the interface deviation amplitude is reproduced periodically in the form of modulations. We point out that the force magnitudes $b_c=1,\, 3$ belong to the  limit of weak instability with the instability parameter $2\alpha/\Omega = 0.095,\, 0.29$, respectively.
Thus, in agreement with the expectations of the analytical model,  the parametric resonance is stabilized at the nonlinear stage in the case of weak instability.
Numerical simulations indicate that the oscillation amplitude increases with increasing magnitude of the driving force as shown in Fig. 5: comparison of the numerical results with Eq. (\ref{maximal-amp}) demonstrates good agreement and allows evaluating  the factor $\kappa$ for the chosen problem parameters as $\kappa = 0.96$.

Taking larger force magnitude $b_c=6$ we obtain the instability parameter comparable to unity, $2\alpha/\Omega = 0.57$, which leads to considerable nonlinear effects in the interface dynamics presented in Fig. 6 for selected time instants $t = 0.08 - 3.46$. Starting with small initial perturbation, the interface oscillates with noticeably smaller frequency than  $\Omega/2$. Besides, the oscillations are accompanied by strong broadening of the interface on the reversal motion of humps transformed into hollows, e.g. at $t= 1.14$. The second growth of the hump on the snapshots $t = 1.66 - 2.08$ is characterized by a complicated shape of the interface, quite distinct from the original harmonic perturbation. We mark resemblance of the interface shape at $t=2.08$ and the "mushroom" structures for the RT instability \cite{Sasaki-2009-RT,Kobyakov-2011-linear}. Due to the large amplitudes  we observe the tendency of   the mushroom cap to detach from the main bulk of the BEC at $t = 2.4$ and to form   a large blob, i.e. a 2D soliton, according to the mechanism discussed in \cite{Bezett-et-al-RM}. The mechanism of the blob detachment is also related to the capillary instability studied in \cite{Sasaki-2011-capillary}. Collapse of the mushroom structure may be also compared to the parametric resonance observed in the combustion system of a flame front in the effective gravitational field produced by sound \cite{Petchenko2006,Petchenko2007}.
Still, in the case of $b_c=6$, detachment of the blob is not complete, and it is followed by reattachment at $t=2.66$. Because of the detachment-reattachment process, the definition of the oscillation amplitude becomes ambiguous, which leads, in particular, to irregularities in the plot of $\zeta(t)$. Reattachment of the blob to the bulk of BEC is accompanied by momentum transfer, i.e. bounce, which produces a grey soliton traveling in the bulk of BEC to the left in the snapshot $t=3.46$. The soliton is formed because mean squared deviation of the external force is well above the critical value $b_{cr}={{\gamma R_0}/{2}}$
discussed in detail in \cite{Kobyakov_QuantumSwapping-2012}. At the same time, interface dynamics for $b_c=6$ does not lead to production of vortices and most of the effects, except production of grey solitons, may be described within the quasi-classical approach.

Generation of vortices may be observed at larger magnitude of the driving force, $b_c=8-10$ in our simulations. The approximate boundary between two regimes is indicated in Fig. 5 by the dashed vertical line, and it corresponds to the instability parameter about $2\alpha/\Omega = 0.8$, i.e. close to unity, in agreement with the expectation of the analytical model. Figure 7 presents amplitude of the interface deviation from equilibrium for $b_c=8-10$ until the instants of droplet detachment and formation of quantum vortices in BEC 1 with the vortex core filled by BEC 2 and vice versa - skyrmions.
After that instant, definition of the interface deviation becomes ambiguous. As we can see in Fig. 7, skyrmions are generated after several interface oscillations for $b_c=8, \, 9$, and on the very first oscillation for $b_c = 10$. Similar to Fig. 6, the oscillations are extremely irregular with characteristic frequency noticeably smaller than $\Omega / 2$. Taking the simulation run for $b_c=8$ as an example, we  observe the front shape in the oscillations for $t < 4.5$ resembling qualitatively the structures presented in Fig. 6. For this reason, we focus on the snapshots for $b_c=8$ and $t\geq 4.5$ in Fig. 8 when qualitatively new quantum effects come to play. In particular, Fig. 8 demonstrates density and phase distributions of BEC 1 for the time instants $t =4.8$ illustrating detachment of the first droplets (2D solitons); $t=4.94$ when the soliton splits into the vortex-antivortex pair; $t=5.12$ when the vortex-antivortex get almost coalescent  because of mutual attraction; $t=7.7$ when the vortex-antivortex are separated again and drifted away from the main bulk of BEC2. In addition, new  vortex-antivortex pairs may be observed at $t=7.7$.

\begin{figure}
\includegraphics[width=3.6in,height=2.2in]{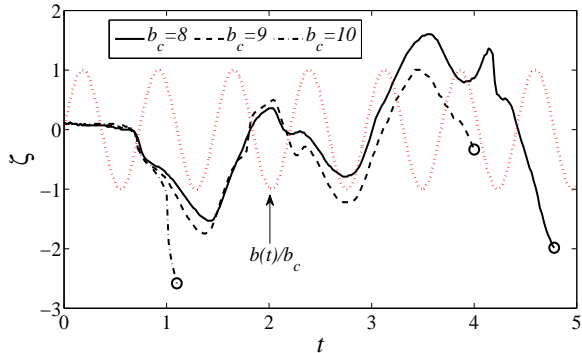}
\caption{(Color online) Amplitude of the interface deviation from  equilibrium  for $R_0=30$ and $b_c= 8 - 10$, and the scaled amplitude of the driving force. The time instants of vortex detachment are indicated by circles. Dimensionless units are defined in Sec II.}
\end{figure}

\begin{figure}
\includegraphics[width=3.6in]{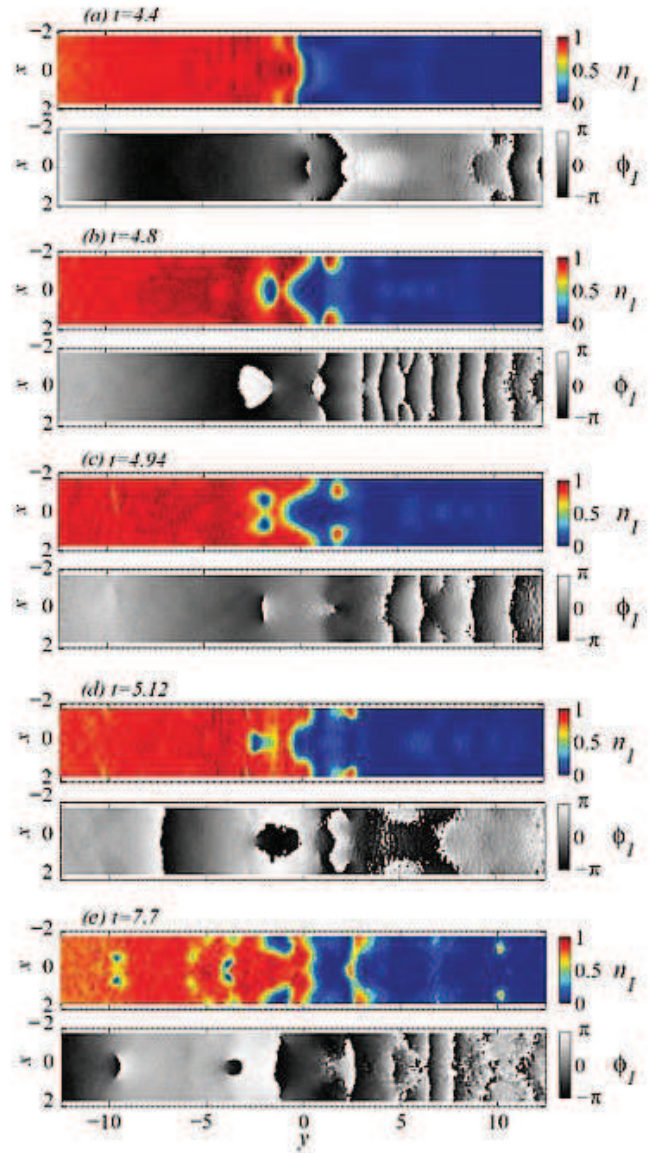}
\caption{(Color online) Snapshots of density distribution of BEC 1, $n_1$, and phase distribution for $R_0=30$ and $b_c=8$, showing vortex generation from the interface between BECs. The detached droplet on (b) splits into a vortex-antivortex pair on (c), and this happens periodically on (d,e) and further, due to periodic external force. Phase plots show topological excitations. Dimensionless units are defined in Sec II.}
\end{figure}

\section{Generation of vortices in the parametric instability}

\subsection{Pseudo-spin description of the two-component BEC}
The regime of large-amplitude force is characterized by generation of quantum vortex-antivortex pairs, so that each of the vortices carries circulation, while total circulation is zero. In the 3D case,  similar vortices would be ring-shaped, though the Kelvin waves can also split the vortex lines into smaller pieces.  The ground state and the vortex structure in the two-component BEC is determined by the interaction parameters. Since the system in study is characterized by weak repulsion, $\gamma\equiv g_{12}/g-1\ll1$, the topological excitations are vortices in one component with the core filled by atoms of another component; such a structure is known as an Anderson-Toulouse (AT) vortex.
The dynamic states shown in Fig. 8 (c, e) present clear examples of the AT vortex pairs; Fig. 9 shows a cut of BEC densities $n_{1,2}$ taken from Fig. 8 (e) at $y=-9.635$, along the line crossing the vortex axes. The obtained vortices may be also described as \emph{skyrmions}, though the term skyrmion refers to a rather wide class of topological solitons. Historically this term originates from the theory by T. H. R. Skyrme \cite{SkyrmeTheory-1961}, which  is quite similar mathematically to description of multicomponent BECs \cite{Topological_obj_in_2compBEC_Cho_2005}. For a small repulsion parameter $\gamma\ll1$, radius of the AT vortex core is given by $\xi_0/\sqrt{\gamma}$, where $\xi_0=\hbar/\sqrt{2mgn}$, \emph{i.e.} it is $1/\sqrt{\gamma}$ times larger than the core radius in a single-component BEC. This value is consistent with the vortex structure in Fig. 9, and it is also comparable to the effective width of the interface between the components \cite{Kobyakov-2011-linear}.

\begin{figure}
\includegraphics[width=3.6in]{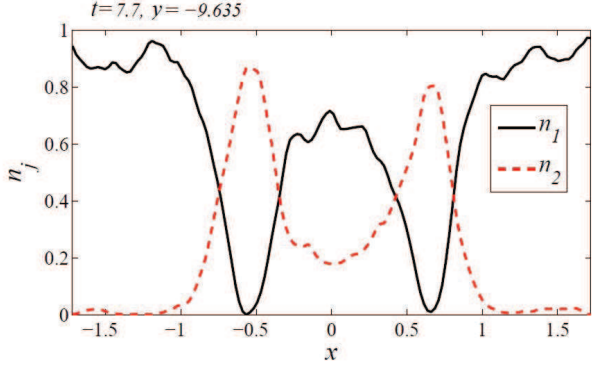}
\caption{(Color online) Cut of $n_1$ and $n_2$ showing dynamic AT-vortex and antivortex of the 1 component (a 2D skyrmion pair) in Fig. 8(d) at t=7.7 and z=-9.635. The 2 component fills the vortex cores of the 1 component. Dimensionless units are defined in Sec II.}
\end{figure}

Let us describe generation of vorticity in our system. A natural way to reveal dynamical difference between single- and multi-component BECs is to go over from the representation of the GP Lagrangian and the wave functions $(\psi_1,\quad\psi_2)$ to the representation of a normalized spinor
\begin{equation}
\mathbf{s}=[\chi_1 \exp({i\beta/2}),\quad\chi_2 \exp({-i\beta/2})]
\end{equation}
and the variables $\sqrt{\rho_T}\exp(i\alpha)$ corresponding to total density $\rho_T$ and total phase $\alpha$. The normalization implies $\chi_1^2+\chi_2^2=1$, and therefore we can choose $\chi_1=\cos{(\chi/2)}$, $\chi_2=\sin{(\chi/2)}$; the new variables $\rho_T,\,\alpha,\,\beta,\,\chi$ are real-valued functions of coordinates and time. If the wave functions are presented as $\psi_j=\sqrt{n_j}\exp (i\phi_j)$, then $\rho_T=n_1+n_2$, $\alpha=(\phi_1+\phi_2)/2$, $\beta=\phi_1-\phi_2$, and $\chi=2\arctan{\sqrt{n_2/n_1}}$.

It is also convenient to define the pseudo-spin vector as $\mathbf{S}=\overline{\mathbf{s}}\mathbf{\sigma}\mathbf{s}$, where $\mathbf{\sigma}=(\sigma_x\quad\sigma_y\quad\sigma_z)^T$, and $\sigma_{x,y,z}$ are the Pauli matrices. Then one readily obtains $S_x=\cos{\beta}\sin{\chi}$, $S_y=\sin{\beta}\sin{\chi}$, $S_z=\cos{\chi}$, with the relation $S_x^2+S_y^2+S_z^2=1$. The superfluid mass current $\rho_T\mathbf{v}_{eff}$ is derived from equations of motion, Eqs. (\ref{GP1}),(\ref{GP2}), as
\begin{eqnarray}
\rho_T\mathbf{v}_{eff}\equiv|\psi_1|^2\mathbf{\nabla}\arg(\psi_1)+|\psi_2|^2\mathbf{\nabla}\arg(\psi_2) \nonumber \\ =\rho_T\left[\mathbf{\nabla}\alpha+{\mathbf{\nabla}\beta\over2}\cos{\chi}\right] \nonumber \\
=\rho_T\left[\mathbf{\nabla}\alpha+{{S}_{z}}\frac{{{S}_{y}}\nabla {{S}_{x}}-{{S}_{x}}\nabla {{S}_{y}}}{2\left( S_{x}^{2}+S_{y}^{2} \right)}\right],
\label{mass_current}
\end{eqnarray}
see \cite{Spin_textures_in_2-comp_BEC_2005} for derivation of the last term. Equation (\ref{mass_current}) demonstrates that vorticity $\nabla\times \mathbf{v}_{eff}$ of the effective superfluid velocity can be non-zero in a multicomponent system without singular regions of the order parameter, in contrast to the single-component case. Using Eq. (\ref{mass_current}) we compute evolution of vorticity in our system, and plot $\nabla\times \mathbf{v}_{eff}$ in Fig. 10 for the same moments of time and the same parameters as shown in Fig. 8. This function may be also interpreted as density of the topological charge, and when integrated over space it gives the topological invariant. This invariant, or Pontryagin index, classifies the stationary states of multi-component BECs; in our system it is formed from the $z$-component of the vorticity pseudo-vector, which can be derived from Eq. (\ref{mass_current}) as \cite{Spin_textures_slowly_rotating_BEC_2004}
\begin{equation}
\label{vorticity}
q(\mathbf{r})={1\over{8\pi}}\epsilon^{ij}\mathbf{S}\left[\partial_i\mathbf{S}\times\partial_j\mathbf{S}\right]={1\over{2\pi}}(\nabla\times \mathbf{v}_{eff})_z.
\end{equation}
In a more general case the proof of Eq. (\ref{vorticity}) can be obtained by considering the form $\chi_a\chi_b$, where $a,\,b$ count the spinor degrees of freedom \cite{Spin_textures_slowly_rotating_BEC_2004}. Topological charge of a single localized AT vortex is $Q=\int d^2\mathbf{r}\,q(\mathbf{r})=1$.

One observes that vorticity $\nabla\times \mathbf{v}_{eff}$ is generated on the interface between BECs. In our superfluid system the interface region may behave locally like a rotating solid body during some finite time intervals. In particular, such behavior may be observed in Fig. 10 (a) for $t=4.4$ , which precedes the detachment of two vortex-antivortex pairs, Fig. 10 (b). When the pairs detach, the interface acquires an opposite effective vorticity sign, Fig. 10 (c), and then, after a period of capillary oscillations, another pair of vortex-antivortex  is generated, Fig. 10 (d). In Fig. 10 we also observe  excitation of compressible modes, which  may lead to dissipations due to two-body processes. However, the dissipation processes are beyond the scope of the GP model and are not considered in the present work.
\begin{figure}
\includegraphics[width=3.6in]{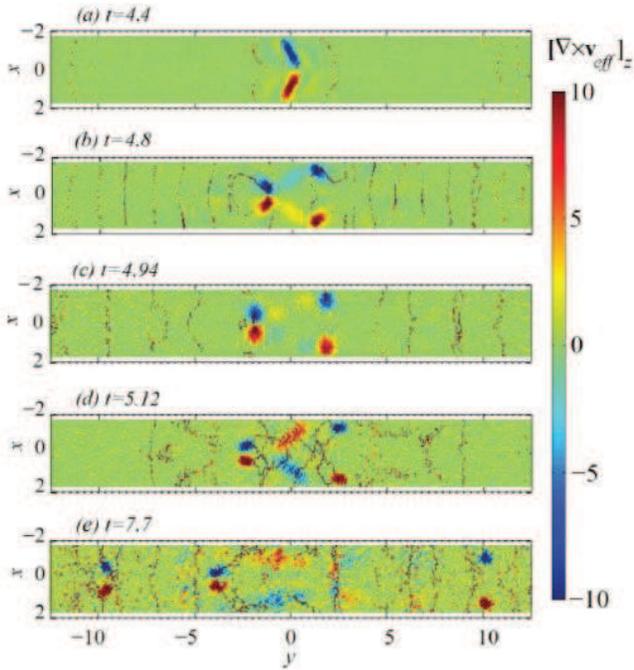}
\caption{(Color online) Snapshots of topological charge density of the two-component BEC for $R_0=30$ and $b_c=8$, for time instants the same as in Fig. 8. Topological charge of the system is generated on the interface between BECs. The direction of the $z$-axis is along $\mathbf{\hat{x}}\times\mathbf{\hat{y}}$. Dimensionless units are defined in Sec II.}
\end{figure}

\subsection{Dynamics of vortices}
The vortex-antivortex pairs are formed from the bubbles, and propagate towards the BEC edge. The pairs are generated on the interface one after another, and they annihilate close to the BEC edge after some sporadic dynamics. In this subsection we investigate dynamics of the first pair on the stage of propagation towards the edge. In our simulations interaction between different vortex pairs is negligible, and a typical trajectory of a pair is represented in Fig. 11, where we glue together cuts from the full density profiles taken at 10 different time moments. The snapshots are limited by the dashed borders which show local density.
\begin{figure}
\includegraphics[width=3.6in]{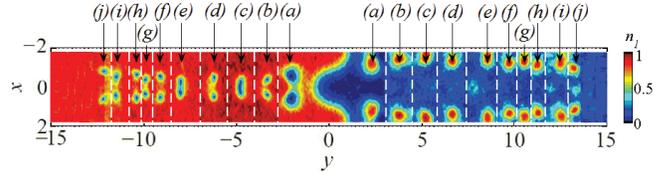}
\caption{(Color online) Snapshots of location of the first generated vortex-antivortex pair in the 1 component (at $y<0$), with corresponding fillings of the cores of the first vortex-antivortex pair of the 2 component (at $y>0$), for $R_0=30$ and $b_c=8$. Snapshots are taken at time instants (a) t = 5.0, (b) 5.4, (c) 5.8, (d) 6.2, (e) 6.6, (f) 7.0, (g) 7.4, (h) 7.8, (i) 8.2, and (j) 8.6, and only the area close the pair is shown for each pair (marked by dashed borders). Note that the time instant $t=5.12$ from Fig. 8(d) is not shown in order to make the picture easier for understanding. The second pairs, generated after the first ones, are not shown for the same reason. At the time instant $t=8.6$, shown on (j), dynamics of vortices is sporadic due to localization of the system and strong influence of the external field. Dimensionless units are defined in Sec II.}
\end{figure}

First let us explain how vortex pairs are formed from the bubbles. Because of the non-linearity of the process of bubble detachment from the interface it is not clear whether one could predict the initial velocity of the bubble $v_0$ immediately after the detachment. However, instead we can predict the \emph{critical} velocity of the bubble, above which it splits into vortex pairs. This velocity results from topology of the order parameter, because quantized vortices are topological excitations. In the TF approximation, the bubble of the BEC 2 produces a cavity in the bulk of the BEC 1, which travels with the bubble velocity, and given that the bulk of BEC 1 is at rest, then, up to logarithmic accuracy, its phase must have a  "jump" (discontinuity) at the cavity center. This jump can exceed $2\pi$, and therefore give rise to formation of the vortices, only when diameter $L$ of the bubble exceeds $2\pi\hbar/mv_0$. Therefore the critical bubble velocity  of the BEC 2 above which the bubble splits into a vortex-antivortex pair, is
\begin{equation}
\label{v_cr}
v_{cr}={{2\pi\hbar}\over{m_1L}},
\end{equation}
where $m_1$ is mass of atom of the BEC 1 (in our numerical simulations we take $m_1=m_2=m$). Equation (\ref{v_cr}) is valid both for  2D and 3D geometries.
We point out that our argument, leading to Eq. (\ref{v_cr}), is based on topology of the order parameter and therefore is general. The only assumption made is validity of the TF approximation $L\gtrsim2\Delta_{int}$. The result Eq. (\ref{v_cr}) can be applied as well for a single component BEC when it is stirred by an obstacle with the size much larger than the healing length of the BEC.
As a result, the number of pairs generated by a bubble with a given initial velocity $v_0$ is equal to the integer part of $v_0/v_{cr}$.
In our system for $R_0=30$, $b_c=8$ we observe formation of a single pair from each bubble.
We measure velocity of the first pair as $\approx1.25\sqrt{\hbar\omega_y/m}$, which is slightly higher than predictions of Eq. (\ref{v_cr}) computed with $L=2\Delta_{int}$ as $v_{cr}\approx0.95\sqrt{\hbar\omega_y/m}$.
In general one should take into account compressibility of the bubble \cite{bubbles_in_2compBECs_2011};  the capillary instability \cite{Sasaki-2011-capillary} for large bubbles may influence the results too.

Equation (\ref{v_cr}) can be used to demonstrate that the lowest critical velocity that breaks the Landau criterion of superfluidity, is related to topological excitations rather than to sound-like excitations. Indeed, the sound speed $c_s$ in the trap center is equal to $\sqrt{2}R_0$, in the dimensionless variables specified in Sec. II. It should be compared with the dimensionless Eq. (\ref{v_cr}), $4\pi/L$, where $L$ is size of the obstacle in units of $a_y$. Using $L=2/(\sqrt{\gamma}R_0)$ with $R_0=30$, and $\gamma=0.01$, relevant for droplets seen in evolution of our system, we obtain $v_{cr}\approx0.44 c_s$, and this is confirmed by the observation that the detached droplets split into vortex-antivortex pairs.

Now we turn to the trajectories of the pairs, which are essentially straight lines modulated by relatively weak oscillations. The straight lines indicate  the stationary propagation of the pair. Indeed, an isolated stationary vortex pair (vortex ring in 3D) moves with a constant velocity with respect to the quiescent bulk when the attraction Magnus force between the vortices in the pair is compensated by the repulsion Magnus force due to the vortex motion. The net Magnus force is equal zero when velocity of the vortex pair is the same as the velocity created by one vortex of the pair in the position of the other. The system is invariant with respect to the Galilean coordinate transformation as can be readily shown in the GP model. However when the atomic cloud, containing a vortex, is subjected to acceleration, the vortex exhibits inertia due to mass of the core filling
\begin{equation}
\label{filling_mass}
m_{f}=m_2 n \pi \Delta_{int}^2,
\end{equation}
where $\Delta_{int}=R_0/\sqrt{\gamma}$ is the  radius of the vortex core, and we have chosen to consider a vortex in the 1 component filled by the 2 component.
On the other hand, the effective mass of the vortex itself may  be obtained taking into account the relativistic character of the collective excitations of BEC coupled to the vortex state \cite{Effective_vortex_dynamics_Wexler_Touless_1996,    Vortex_mass_Tunneling_Volovik_1997}:
\begin{equation}
\label{vortex_mass}
m_v=E_v/c_s^2,
\end{equation}
where $E_v=\pi n_1 \hbar^2 \ln\left({R_{tot}/\Delta_{int}}\right)/m_1$ is energy of a vortex in the BEC 1  in a cylindrical system limited by a container with a large radius $R_{tot}$, and $c_s=\sqrt{gn/m}$ is the sound speed in the trap center. Thus the effective action, describing the degree of freedom of the system associated with the position of the vortex line element $\mathbf{R}(t)$, contains the "kinetic" contribution of the form
\begin{equation}
\label{S_kin}
S_{kin}=\int dt \, (m_v+m_f) {\dot{\mathbf{R}}}^2/2.
\end{equation}
The "potential" contribution describes forces acting on a vortex. It includes the Stern-Gerlach force applied to the filling composed of atoms of the 2 component (in our system the force is directed along $y$)
\begin{equation}
\label{S_ext}
S_{ext}=-\int dt\, {m_{f}\over m_2}R_y {\mu_B B'\over 2},
\end{equation}
and the lift force, i.e. the quantum analogue of the Magnus force, which may be derived from the GP model. In the derivation we start from the action for a two-component BEC
\begin{equation}
\label{action}
S=\int{dt}\int{d^{3}\mathbf{r}}\left[\sum_{j=1,2}\textrm{Re}\left(i\hbar\psi_j^*\partial_t\psi_j\right)-\mathcal{E}\right].
\end{equation}
The "potential" contribution to the action associated with interaction of the vortex with the background velocity is
\begin{eqnarray}
&&S_{M}=\int dt d^2\mathbf{r}\,\textrm{Re}[i\psi_1^*(\mathbf{r}-\mathbf{R}(t))\partial_t\psi_1(\mathbf{r}
-\mathbf{R}(t))]= \nonumber \\
&&{i\over 2}\int dt d^2\mathbf{r}\,\left[\psi_1^*(\mathbf{r}-\mathbf{R}(t))\mathbf{\nabla}\psi_1(\mathbf{r}-\mathbf{R}(t))-
\nonumber\right. \\
&&\left.\psi_1(\mathbf{r}-\mathbf{R}(t))\mathbf{\nabla}\psi_1^*(\mathbf{r}-\mathbf{R}(t))\right]\mathbf{\dot{R}},
\end{eqnarray}
where $\psi_1$ is the stationary wave function of a vortex (the chemical potential was dropped for brevity because it does not alter the result). The wave function $\psi_1(\mathbf{r}-\mathbf{R}(t))$ is Taylor-expanded for small $\mathbf{R}(t)$ as
\begin{equation}
\psi_1(\mathbf{r}-\mathbf{R}(t))=\psi_1(\mathbf{r})+\mathbf{R}(t)\nabla\psi_1(\mathbf{r})+O(\mathbf{R}^2,...),
\end{equation}
and only the first order in $\mathbf{R}(t)$ is kept. After that the spatial integration is readily done, and we obtain for a singly-quantized vortex in the 1 component
\begin{equation}
\label{S_M}
S_{M}=-\nu_1\pi\hbar n_1 \int dt \epsilon^{ij}R_i \dot{R}_j,
\end{equation}
with $\epsilon^{12}=1$, $\epsilon^{ij}=-\epsilon^{ji}$, $i,j=x,y$, and $\nu_1=\pm1$ is the winding number of the vortex with the effective vorticity along $z$ ($+1$), or in the opposite direction ($-1$). It follows from the energetic conditions, and from our real-time numerical simulations, that vortices with $|\nu_1|>1$ are not formed  during the evolution of our system. Thus, vortices and anti-vortices experience opposite forces due to a background superfluid flow with velocity $\mathbf{\dot{R}}$. As a result, a single AT vortex in a homogenous two-component BEC is described in 2D by the action
\begin{equation}
\label{vortex_action}
S_v[\mathbf{R},\mathbf{\dot{R}}]=S_{ext}+\int dt \left[(m_v+m_f) {\dot{\mathbf{R}}}^2/2 -\nu_1 \pi\hbar n \epsilon^{ij}R_i \dot{R}_j    \right].
\end{equation}
We note that the action of the external force on the 1 BEC is not explicit in Eq. (\ref{vortex_action}), because it implicitly contributes to ${\dot{\mathbf{R}}}$ by inducing collective bulk oscillations.

Equation (\ref{vortex_action}) describes the quantum analogue of the Magnus force, which acts perpendicularly to velocity of a vortex $\mathbf{\dot{R}}(t)$. Using Eq. (\ref{vortex_action}) and going to dimensionless variables specified in Sec. II, we obtain  the dimensionless form of the force acting on a vortex in the 1 BEC with non-singular core filled by the 2 BEC
\begin{equation}
\label{magnus_dimensionless}
\partial^2_{tt}R_i=-{4\nu_1{\gamma R_0}\epsilon^{ij}\dot{R}_j+b_i(t)\over{1+2\gamma\ln{(\gamma R_{tot}/\xi_0)}}},
\end{equation}
where $b_i(t)$ are Cartesian components of the external force applied to the 2 BEC, $b_y(t)=b(t)$, see Eq. (\ref{eq1.02}). Equation (\ref{magnus_dimensionless}) shows that for systems with size $R_{tot}\lesssim(2/\sqrt{\gamma R_0^2})\exp({{1/{2\gamma}}})$ the vortex mass is dominated by mass of the core filling. In our system $R_{tot}\approx R_0$, and the vortex mass is approximately that of the filling.
We conclude that the oscillations of the pair size observed in Fig. 11, are due to the bulk motion driven by the external force, and due to the external force applied to the filling. Motion of the bulk determines bulk velocity with respect to vortices, and hence the Magnus force. As follows from Eq. (\ref{magnus_dimensionless}), the frequency of the oscillations  is close to the frequency of the external force, i.e.  $\approx\Omega=8.55$, which is confirmed by the numerical solution shown in Fig. 11.

\section{SUMMARY}
We have studied the parametric instability at the interface of two immiscible BEC components pushed towards each other by an oscillating force. The instability develops due to the parametric resonance, which pumps quantum capillary waves at the interface. At moderate amplitudes of the driving force the instability is stabilized at the nonlinear stage due to modifications of the doubled oscillation frequency in comparison with frequency of the driving force.
In that case the BEC interface demonstrates oscillations with modulated amplitude, which depends on the strength of the driving force.
When the amplitude of the driving force is large enough, we observe  detachment of droplets from BEC components and generation of quantum vortices -- skyrmions. The skyrmions are born as vortex pairs and move almost stationary from the interface to the trap edge. The properties and dynamics of the skyrmion pairs are discussed. We have introduced the critical velocity $v_{cr}$ of a droplet of one BEC in a bulk of another BEC from topological arguments, and derived the analytical formula for the quantum counterpart of the Magnus force acting on skyrmions.

\emph{Note added in the proof.} Recently, the authors became aware
of the work \cite{Balaz_Nicolin_parametric_2011} on the Faraday resonance in two-component
BEC confined to a quasi 1D geometry. The main difference
between the current work and Ref. \cite{Balaz_Nicolin_parametric_2011} is in the geometry
(multidimensional versus 1D, respectively) and in the nature
of the surface waves amplified by the oscillating external force.
Here we study interfacial capillary modes on the common
surface of two BECs, while in the work \cite{Balaz_Nicolin_parametric_2011} these are the
surface modes at the edges of each BEC facing vacuum.

\acknowledgements
This research was supported partly by the Swedish Research Council (VR) and by the Kempe Foundation. Calculations have been conducted using the resources of High Performance Computing Center North (HPC2N). D.K. is grateful to Professor C. J. Pethick for inspiring discussions.

\end{document}